 \documentclass[final,3p,times,twocolumn]{elsarticle}
\usepackage{amssymb}
\begin{document}

\begin{frontmatter}



\title{Photometric study of five open star clusters}


\author{Sneh Lata\footnote{E-mail: sneh@aries.res.in (S. Lata)}, A. K. Pandey$^1$, Saurabh Sharma$^1$, Charles Bonatto$^2$ and Ram Kesh Yadav$^1$}

\address{$^1$Aryabhatta Research Institute of Observational Sciences, Manora Peak, Nainital 263002, Uttarakhand, India\\
        $^2$Universidade Federal do Rio Grande do Sul, Departamento de Astronomia CP 15051, RS, Porto Alegre 91501-970 Brazil\\
}

\begin{abstract}
$UBVRI$ photometry of the five open clusters Czernik 4, Berkeley 7, NGC 2236,
NGC 7226 and King 12 has been carried out using ARIES 104 cm telescope, Nainital. Fundamental
cluster parameters such as foreground reddening $E(B-V)$, distance, and age have been
derived by means of the observed two colour and colour-magnitude diagrams, coupled
to comparisons with theoretical models. $E(B-V)$ values range from 0.55 to 0.74 mag,
while ages derived for these clusters range from $\sim$10 to $\sim$500 Myr. We have also
studied the spatial structure, mass function and mass segregation effects. The present
study shows that evaporation of low mass stars from the halo of the clusters increases
as they evolve.
\end{abstract}


\end{frontmatter}

\section{Introduction}
Open clusters (OCs) are excellent tools to understand star formation and evolution
because their stars emerge from the same molecular cloud and have a mass spectrum that can be used 
to study the initial mass function (IMF).
Poorly populated OCs do not survive longer than a few hundred Myr  
, whereas rich ones may survive longer (e.g. Pandey \& Mahra 1986;
 Theuns 1992;
Tanikawa \& Fukushige 2005; Carraro et al. 2005; Bonatto et al. 2012). OCs are also one of the best tools
to probe the age and abundance structure of the Galactic disk. As clusters evolve
through dynamical effects (internally and by interactions with the Galactic tidal
field), a significant fraction of the cluster stars is gradually lost to the field.
Mass segregation and evaporation of low-mass members are the main effects of dynamical
interactions (de la Fuente Macros 2001; Andersen \& Nordstr{\"o}m 2000; Patat \& Carraro 1995; Carraro 2006). Hence, to understand cluster evolution it is necessary to study
the dense central region (the core) as well as the expanded and sparse region
(the halo or corona) (Pandey et al. 1988; Maciejewski 2009).
In addition, study of the structure of OCs also gives the opportunity to understand
how external environments and internal stellar encounters affect OCs. The
morphological structure, or shape, of the young OCs can be
governed by initial conditions in the molecular clouds and by internal gravitational
interactions and external tidal perturbations as the cluster evolves (Chen et al. 2004; Sharma et al. 2006, 2008). Based on N-body
simulations, de la Fuente Marcos (1997) finds that the total disruption time for a cluster
also depends on its richness.
 The studies of dynamical properties of the OCs are difficult in the absence of kinematical data. 
However, some information about the dynamical
evolution of open clusters can be drawn from statistical analysis of the spatial distribution
of the probable cluster members and their mass function (Kang \& Ann 2002; Ann \& Lee 2002).

In order to continue our efforts to study the dynamical evolution of OCs using 
the photometric properties, we carried out photometric observations of OCs Czernik 4 (Cz 4), Berkeley 7 (Be 7), 
NGC 2236, NGC 7226 and King 12, which cover a wide range in age and are relatively poorly studied.
The basic parameters of these clusters available in the literature are listed in Table 1.
Additionally, since these clusters are of intermediate to old age, they may present
observable consequences of mass segregation. 
   
The paper is organized as follows. We describe observations and data reduction technique in the Section 2. The Section 3 deals with spatial structure of the
cluster. The Section 4 describes basic parameters of clusters using the two colour diagram and colour-magnitude diagram obtained in the present study. We study luminosity and mass function in section 5.
In Section 6 we put our effort in describing dynamical evolution of clusters by means of their MF and structural parameters.
Finally, we summarize our results in Section 7.

\section{Observations and Data Reduction}
The $UBVRI$ CCD observations of open clusters  Cz 4,  Be 7, NGC 2236, NGC 7226 and King 12 were carried out using the 104 cm ARIES Telescope. A 2k$\times$2k
 CCD was used as a detector. The field of view is $\sim$13$^{\prime}\times$13$^{\prime}$
and the plate scale is $\sim0.76^{\prime\prime}$/pixel in 2$\times$2 pixel binning mode.
The log of the observations is given in Table 2.
The observed $V$-band images of the clusters are shown in Fig. 1. 
The typical seeing (estimated from the FWHM of the point spread function; PSF) of the images was found to be $1.5^{\prime\prime}-2^{\prime\prime}$. Bias and twilight flats were also taken
along with the target field. 
The preprocessing of the CCD images was performed by using the IRAF\footnote{IRAF is distributed by the National Optical Astronomy Observatory, which is operated by the Association of Universities for Research in Astronomy (AURA) under cooperative agreement with the National Science Foundation.},
which includes bias subtraction, flat field correction and removal of cosmic rays.
The instrumental magnitude of the stars were obtained using the DAOPHOT package provided by Stetson (1987, 1992). Both aperture and PSF photometry were carried out to get the magnitudes of the
stars. The PSF photometry yields better results for crowded regions.
The standardization of the cluster fields was carried out by observing the standard field SA 98 (Landolt 1992).  Instrumental magnitudes were converted to standard magnitudes using the following transformation equations

\noindent $v = V + q{_1}+p_{1}(V-I)+k_{v}X$ \\
$b = B + q{_2}+p_{2}(B-V)+k_{b}X$ \\
$i = I + q{_3}+p_{3}(V-I)+k_{i}X$ \\
$r = R + q{_4}+p_{4}(V-R)+k_{r}X$ \\
$u = U + q{_5}+p_{5}(U-B)+k_{u}X$ \\

\noindent In  the above  equations $u$, $b$,  $v$, $r$
and $i$, obtained after time and aperture corrections are the  instrumental magnitudes
while $U$, $B$, $V$, $R$ and $I$ are the standard magnitudes; $p_{1}$,
$p_{2}$,  $p_{3}$, $p_{4}$  and $p_{5}$  are the  colour coefficients;
$q_{1}$,  $q_{2}$, $q_{3}$,  $q_{4}$ and  $q_{5}$ are  the  zero point
constants;  $k_{v}$, $k_{b}$,  $k_{u}$,  $k_{r}$ and  $k_{i}$ are  the
extinction coefficients in $V$, $B$,  $U$, $R$ and $I$ filters, respectively, and $X$
is the airmass. The values of the coefficients 
 for each observing  night are listed in Table  3. 
The  typical DAOPHOT  errors at brighter level ($V\approx$ 16 mag) are $\le$ 0.01 mag, whereas for fainter end ($V\approx$ 20 mag) the errors become larger ($\approx$ 0.05 mag).
The entire $UBVRI$ CCD photometric  data for the cluster under present study is  available in electronic  form
at  the  WEBDA  open cluster  data  base  website  at
http://obswww.unige.ch/webda/.   It  can  also  be obtained  from  the
authors.

\subsection{Comparison with previous photometry}
Barring the cluster Cz 4 all the other clusters have already been studied partially either photoelectrically or photographically. 
A comparison of the present photometries for clusters Be 7, NGC 2236, NGC 7226 and King 12 with the photometries available in the literature has been carried out. 
In Fig. 2 we plot $\Delta V$, $\Delta (U-B)$ and $\Delta (B-V)$ (in the sense, present data minus literature data) for the common stars as a function of $V$ magnitude. 
\\
\\
\noindent{\it Be 7}\\
Phelps \& Janes (1994) carried out CCD photometry of Be 7 in the area of 11.68$\times$11.68 arcmin$^2$ for the first time and reported 722 stars down to $V$ $\sim$ 21 mag. 
The comparison shows that the present $V$ magnitude and $(B-V)$ colour are in fair agreement with those obtained by Phelps \& Janes (1994), 
whereas in the case of  $(U-B)$ colours there seems to be a some trend in $\Delta (U-B)$. 
\\
\\
\noindent{\it NGC 2236}\\
The photographic observations for NGC 2236  were 
first carried out by Babu (1991; 26 stars) and Malisheva \& Shashkina (1987; 434 stars).
Babu (1991) also presented $UBV$ photoelectric data for the region containing the cluster NGC 2236.
Comparison has been made between the present data and those available in the literature. In Fig. 2 open circles and trinagles represent present data minus
photographic data given by Babu (1991) and Malisheva \& Shashkina (1987) respectively, whereas filled circles represent present data minus
photoelectric data given by Babu (1991). 
There seems to be a large scatter in the $\Delta V$, $\Delta (B-V)$, $\Delta (U-B)$ without any trend.
However, the $V$ magnitude and $(B-V)$ colour made photoelectrically (see Fig. 2 filled circles) by Babu (1991) seem to match with those obtained in present study. 
\\
\\
\noindent{\it NGC 7226}\\
The CCD photometric observations in an area of 7.9$\times$7.9 arcmin$^2$ were presented by Viskum et al. (1997). They reported photometry for 499 stars down to $V\sim$ 18 mag. 
The present $V$ magnitudes and $(B-V)$ colours are in fair agreement with those obtained by Viskum et al. (1997).
\\
\\
\noindent{\it King 12}\\
Mohan \& Pandey (1984) presented $UBV$ photoelectric photometry for
30 stars in the cluster. The present magnitudes and colours match well with those
reported by Mohan \& Pandey (1984).
\\
\\
 
\subsection {Completeness of the data}
It is not possible to detect all the stars present in the CCD frame, especially towards
the faint tail. The main reason for the non detection of stars could be e.g, faintness
and crowding in the region. The ADDSTAR routine given in the DAOPHOT
software package (Stetson 1987) was used to correct the data
for incompleteness. The artificial stars were added at random positions in the CCD images. The luminosity distribution of artificial stars was chosen in such a way that more
stars are inserted towards the fainter magnitude bins. In
all, about 10\%-15\%
 of the total stars were added so that
the crowding characteristics of the original frame do not
change significantly. The ratio of the stars recovered to those
added in each magnitude interval gives the completeness factor (CF) as a function of magnitude. The frames were re-
reduced in a similar procedure as used in the case of the
original frames. In practice, we followed the procedure given
by Sagar \& Richtler (1991). We added artificial stars to both
$V$ and $I$ images in such a way that they have similar geometrical locations but differ in $I$ brightness according to mean
$(V-I)$ colours of the MS stars. The minimum value of the
CF of the pair thus obtained is used to correct the data for
incompleteness (Sagar \& Richtler 1991). As expected, the
incompleteness increases towards fainter magnitudes. The
data are found to be complete by 90\%
 upto the magnitude of
$V\sim$19 mag. 

\section{Spatial Structure of the clusters: Radial density profile}
The radial density profile has been used to probe the structure and estimate the extent ($r$) of the 
clusters. 
Most of the stars are found to be located in the central
region, and the stellar density decreases with radial distance which finally merges
with the field. The point of maximum stellar density is considered as the
cluster centre. To determine the radial density profile, we
made several concentric rings around the cluster centre. The
stellar density is determined by counting stars inside each
concentric annuli and then dividing the number of stars by
the area of the respective annulus. We plot the stellar density as a
function of radial distance in Fig. 3. The point at which the
stellar density merges with the field star density is considered the cluster extent. The field star density is estimated
from the region well outside the cluster extent. The estimated radii are: 3.5 arcmin for Cz 4, 4.1 arcmin for Be 7,
 4.4 arcmin for NGC 2236, 3.7 arcmin
for NGC 7226 and 4.0 arcmin for King 12. 

The observed radial density profile of each cluster was fitted by the King profile using the relation $\rho(r)=f_{0}/(1+(r/r_{c})^2)$ (Kaluzny \& Udalski 1992) where $\rho (r)$ is  the projected radial density. 
The $r_c$ is the clusters's core radius which is defined as the radial distance at which
the value of $\rho (r)$ becomes half of the central density $f_{0}$.
Fig. 3 shows the best fit radial density profile obtained through a $\chi^{2}$ minimization technique.  
The cluster core radius thus found for the clusters Cz 4, Be 7, NGC 2236, NGC 7226 and King 12  are 1.09$\pm$0.16,
0.54$\pm$0.07,
1.29$\pm$0.15, 0.72$\pm$0.08 and 1.09$\pm$0.10 arcmin
respectively.

\section {Fundamental parameters of the clusters}
The $V/B-V$, $V/V-R$ and $V/V-I$ colour magnitude diagrams (CMDs) for the clusters are shown in Fig. 4, which show a well defined main-sequence (MS), however the contamination due to field stars increases with increase in the magnitude.
Therefore, to estimate the field star contamination we have used
 a region outside the cluster region. The area of the field region was kept equal to the area of the cluster region.
The contribution of field stars from the CMDs of the cluster region was statistically removed using the procedure described by Sandhu et al. (2003) and Pandey et al. (2007). Briefly, for a star in the $V/V-I$ CMD of the field region, the nearest star within $V$$\pm$0.25 and $(V-I)$$\pm$0.125 in the cluster's $V/V-I$ CMD was removed. While removing stars from the cluster CMD, the number of stars in each magnitude bin was maintained as per the CF. The statistically cleaned two colour diagram (TCD) $(U-B/B-V)$ and CMDs ($V/B-V, V/V-I$, and $V/V-R$) for target clusters are shown in Figs. 5, 6, 7 and 8 and used to estimate their fundamental parameters.

\subsection {Reddening}
 The reddening $E(B-V)$ has been determined using the $U-B/B-V$ TCD of the
clusters by comparing the theoretical model of Z=0.02 by Girardi et al. (2002) with the observations assuming the normal reddening law, i.e, $E(U-B)/E(B-V)=0.72$, in the  
regions. The $U-B/B-V$ TCDs for the cluster region are shown in
Fig. 5. 
In case of King 12 we have taken data for three brighter stars from Mohan \& Pandey (1984), which could be probable members of the cluster. These stars were saturated in the present observations of King 12. The fundamental parameters like reddening, distance and age for King 12 have been determined after adding data of these stars. 
The $U-B)$/$(B-V)$ TCD yields  $E(B-V)=$0.55, 0.74, 0.56, 0.60 and 0.58 mag towards the clusters Cz 4, Be 7, NGC 2236, NGC 7226 and King 12, respectively. The estimated uncertainty in estimation of the $E(B-V)$ values is $\sim$ 0.05 mag. The $E(B-V)$
obtained in the present study are in fair agreement with the values available in the literature (0.80 mag for Be 7, by Phelps \& Janes 1994; 0.37 mag for NGC 2236, by Rahim 1970 and 0.68-0.84 mag, by Babu 1991; 0.49 mag and 0.46 mag for NGC 7226, by Yilmaz 1970 and Viskum et al. 1997; 0.58 mag for King 12, by Mohan \& Pandey 1984).
In case of NGC 2236 Babu (1991) found a variable reddening $E(B-V)$ across the field of the cluster ranging between 0.68 mag and 0.84 mag. Mohan \& Pandey (1984) reported non-uniform reddening in case of King 12. The minimum and maximum reddening values estimated by them are 0.52 mag and 0.69 mag. 
\subsection {Distance and age}
The distances and ages of the clusters have been 
estimated by visual fitting of the theoretical isochrones of metallicity Z=0.02 by Girardi et al. 
(2002) with the MS. The $V/B-V$, $V/V-R$ and $V/V-I$ CMDs along with the visually
fitted isochrones are shown in Figs. 6, 7 and 8. The $E(B-V)$ values for each cluster as mentioned in Section 4.1 has been used in fitting the isochrones. The estimated values of distance modulus $(m-M)$ and log (age) for each cluster are given in the Table 4. 
The values for distance and age for each cluster given in the literature are also mentioned in the Table 4.
The expected error in estimation of $(m-M)$ by visual fitting of the theoretical models to the observations is
of the order of 0.1 mag.
The error  on the age  is mainly due  to the visual fitting,  which is
expected to be $\sim$0.1 in the  logarithmic scale of ages except for NGC 7226 (see Table 4). 

The cluster Be 7 is one of the youngest clusters in the present sample.
 Phelps \& Janes (1994) estimated its age as $\sim$4 Myr, whereas in the present work we estimated its age 12.58 Myr. 
The present estimation for reddening, distance modulus and age are in fair agreement with those
given by Phelps \& Janes (1994).   
The cluster NGC 2236 has  well defined MS and significant number of evolved stars. We have determined the distance modulus and log (age) of the cluster as 
14.0 mag and 8.7. This cluster is the oldest cluster in the present sample. 
The values of distance modulus and log (age) listed in WEBDA are 13.82 mag and 8.54 which match with the present observations.
The present distance modulus and age for NGC 2236 also agree with those derived by Rahim (1970) and Babu (1991).  
The present derived log (age) for cluster the NGC 7226 is log (age)=8.4 which is in agreement with the age listed in WEBDA (log (age)=8.4) and the value (log (age)=8.6) derived by Viskum et al. (1997), while present distance estimation is somewhat larger (about twice) than that given in the WEBDA.
The age and distance of cluster King 12 given in Mohan \& Pandey (1984) are based on $UBV$ photoelectric data. 
However, the distance modulus $(m-M)=14.1$ mag obtained in the present study seems comparable with the distance modulus estimated by Mohan \& Pandey (1984). The value of log (age) given by Mohan \& Pandey (1984) matches well with the
present value of log (age)=7.1.

\section{Luminosity and Mass Function}
The MS luminosity function (LF) is estimated by counting number of stars in magnitude bin. The star counts
have been made in 1.0-mag bins in $V$ of the stars lying inside the strip.
The LF can be converted in to mass function (MF) using the theoretical models.
We used the model by Girardi et al. (2002) to convert the LF to MF.
The MS LF for the core, corona and whole
cluster region are derived using the statistically  cleaned $V/V-I$ CMDs and these are given in Table 5.
The obtained MFs for the
core, corona and whole cluster region are shown in
Fig. 9.

The MF generally follows the power law, $N (\log M) \propto M^{\Gamma}$, and the slope of the MF is given as: \\

$\Gamma= d $log$ N ($log$ M)/d $log$ M $ \\

\noindent where  $N ($log$ M)$ denotes the number  of stars in a logarithmic mass bin
and $\Gamma$ is the MF slope.
The classical value derived by
Salpeter (1955) for the slope of IMF in the solar neighbourhood for the mass range $0.4<M/M_{\odot}< 10$ is $\Gamma=-1.35$.
The values of the MF slopes in the core, corona and whole cluster region
are given in Table 6. The slopes of the MF in the clusters Be 7, NGC 2236 and King 12 for whole cluster are found to be comparable with the Salpeter value, whereas in the case of Cz 4 and NGC 7226 the value of slope is found to be shallower and steeper respectively, in comparison to the Salpeter value. However, the values of the slope of MF for
all the clusters are comparable with 3$\sigma$ errors with the Salpeter value.
Barring the case of NGC 2236
the value of $\Gamma$ is found to be steeper in the outer regions of the clusters as compared to that of the core region.
This suggests that a large number of low mass stars may have migrated towards the corona, indicating an effect of mass segregation.

\section{Dynamical Evolution: Mass Segregation}
The cumulative radial distribution  for two mass groups 
 is  shown  in
Fig. 10. Effects of mass segregation are  apparent in all the
clusters under present study.
The conclusion is further checked by using the
Kolmogorov-Smirnov (KS) test.
The confidence level ($\ge$ 95\%) 
confirms that the distribution of low mass stars are segregated from the massive ones, in the sense that massive stars are preferentially located towards the center of the cluster.

To  see whether  the  mass  segregation present in the clusters is  result  of
dynamical evolution,  we have  computed the dynamical  evolution time,
$T_{E}$, using the following relation \\

$T_{E} = \frac{8.9\times10^5\sqrt{N}R_h^{3/2}}{\sqrt{\bar{m}}\log({0.4N})}$ \\

\noindent where  N is  the number of  cluster members, R$_{h}$  is the
radius  within  which  half  of  the cluster  mass  is  contained  and
$\bar{m}$ is  the average mass of  the cluster stars  (Spitzer \& Hart
1971).  The value of R$_{h}$, assumed to be
 equal to half of the cluster extent, 
 is given in  Table 7.
The total number of MS stars in the given mass range (see Table 6)
are obtained with the help of the MF.
A comparison of  the cluster's age with their  dynamical relaxation time
$T_{E}$, given in Table 7,
indicates that the former is larger than the latter. We conclude that
all  the clusters  are dynamically  relaxed and  the observed mass segregation
 seems to have reached some level of dynamical relaxation. 

In order to study the effects of the dynamical evolution on the MFs,
we calculated dynamical evolution parameter for each
cluster $\tau$,
which is defined as, $\tau = age/T_{E}$. Table 7 lists the estimated values of $\tau$ for each cluster.
In Fig. 11.  we plot the $\Gamma$ as a function of $\tau$. To increase the sample we added data from Sharma et al. (2008) and Lata et al. (2010).
Although the scatter is large, Fig. 11a and Fig. 11b show that a systematic decreasing trend in $\Gamma$ with $\tau$
as suggested by Sharma et al. (2008),
particularly in the outer region of the clusters. Bonatto \& Bica (2005) have also
found the same.
However, contrary to the findings of Bonatto \&
Bica (2005) and  Maciejewski \& Niedzielski (2007), we did not find any relation of MF slope for core
region with $\tau$, and the same is concluded by Sharma et al. (2008). The
relation of MF slope with $\tau$ can be represented by following relation; \\
$$\Gamma = \Gamma_{0}+exp(a / \tau), $$
with  $\Gamma_{0}=-1.89\pm0.14$ and $a=-9.15\pm5.41$ for
the whole cluster  region;  $\Gamma_{0}=-2.07\pm0.13$ and $a=-8.77\pm4.10$ for
outer region.
To study the effects of external interactions as well as internal
gravitational interactions on the mass segregation, a quantity $\Delta \Gamma = \Gamma_{core}-\Gamma_{corona}$; is plotted as a function of log (age), $\tau$, and $R_{G}$ in Fig. 12. The Galactocentric distance, $R_{G}$, has been calculated assuming distance between Sun and Galactic center as 8.5 kpc.   
Fig. 12a does not show any relation between $\Delta \Gamma$ and  log (age).
The $\Delta \Gamma$ seems to decrease with  $\tau$ (Fig. 12b) as suggested by Sharma et al. (2008).
The decrease in $\Delta \Gamma$ with $\tau$ can be interpreted as evaporation
of low mass stars from the outer region. Fig. 12c also indicates a systematic
variation of $\Delta \Gamma$ as a function of Galactocentric distance except NGC 2236,
indicating that evaporation of low mass members from outer region of the clusters is not significant at larger Galactocentric distances. 
Cluster NGC 2236 does not follow the trend due to fact that it is the oldest cluster in the present sample and located towards the Galactic center.
This scenario could be interpreted as the dynamical evolution drives the star cluster toward core collapse; the most massive stars segregate to the cluster center while lower mass stars segregate to the cluster halo. At the same time the cluster halo is stripped by the Galactic tidal field.
The low mass stars therefore tend to be lost from the cluster at larger distances from the Galactic center while higher mass stars can reach farther in (McMillan \& Portegies Zwart 2001).
Here we would like to point out that
more sample is needed to get a conclusive view about variation of $\Delta \Gamma$ with log (age), dynamical evolution parameter ($\tau$) and Galactocentric distance ($R_{G}$). 

\section {Summary}
This work presents $UBVRI$ CCD photometry of five open clusters namely Cz 4, Be 7, NGC 2236, NGC 7226 and King 12. Fundamental parameters such as cluster extent, reddening $E(B-V)$ and distance to the cluster have been obtained using the
optical data. The cluster extents are 3.5 arcmin for Cz 4, 4.1 arcmin for Be 7, 4.4 arcmin for NGC 2236, 3.7 arcmin for NGC 7226 and 4.0 arcmin for King 12. 
The values of reddening $E(B-V)$ for the Cz 4, Be 7, NGC 2236, NGC 7226 and King 12 are estimated as 0.55, 0.74, 0.56, 0.60 and 0.58 mag   
respectively. The ages of the clusters have been derived by visually fitting the theoretical models to the observed data points. The ages for present cluster sample range from $\sim$10 Myrs to $\sim$500 Myrs. 
 The mass function slope for whole cluster region of the present cluster sample is found to be comparable to the Salpeter value.
All the clusters show effects of mass segregation. The ages of the clusters are larger than their relaxation times, indicating that all the cluster are dynamically relaxed. The decrease in $\Delta \Gamma (=\Gamma_{core}-\Gamma_{corona})$ with $\tau (= age/T_{E})$ can be interpreted as evaporation
of low mass stars with time from the outer region of the cluster.

\bibliographystyle{model1a-num-names}
\bibliography{<your-bib-database>}
%\section{Thebibliography}

%----------------------------------------------------------------
\begin{figure*}
\vspace{-2.cm}
\hbox{
\hspace{-3.0cm}
\includegraphics[height=8cm,width=12cm]{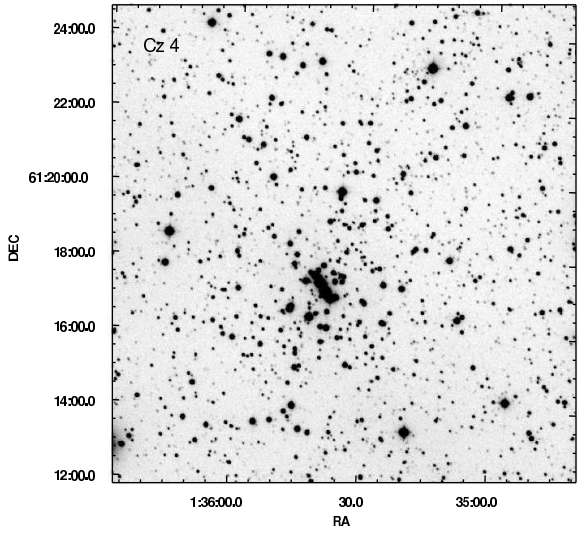}
\hspace{-3.0cm}
\includegraphics[height=8cm,width=12cm]{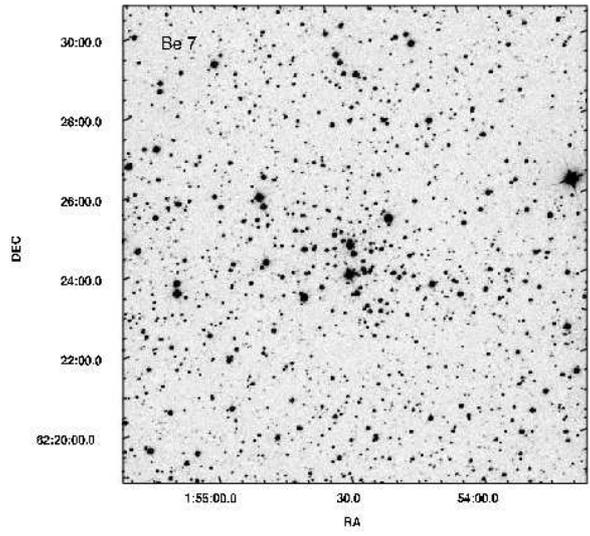}
}
\vspace{-0.2cm}
\hbox{
\hspace{-3.0cm}
\includegraphics[height=8cm,width=12cm]{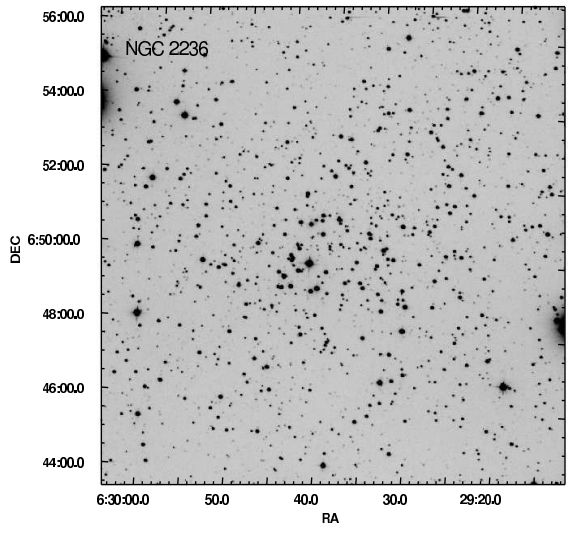}
\hspace{-3.0cm}
\includegraphics[height=8cm,width=12cm]{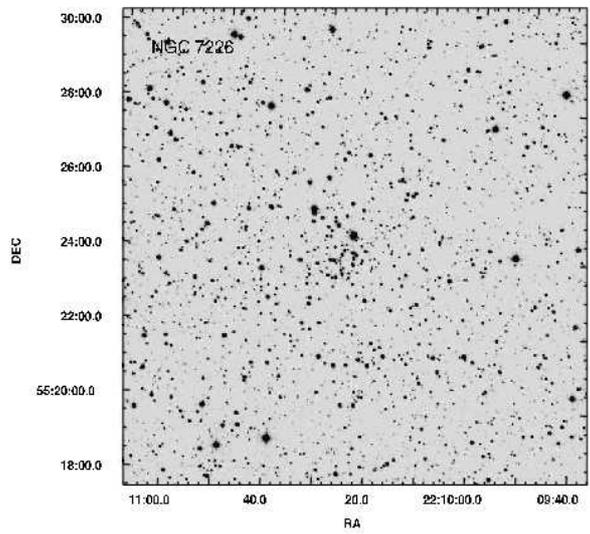}
}
\vspace{-0.2cm}
\hbox{
\hspace{-3.0cm}
\includegraphics[height=8cm,width=12cm]{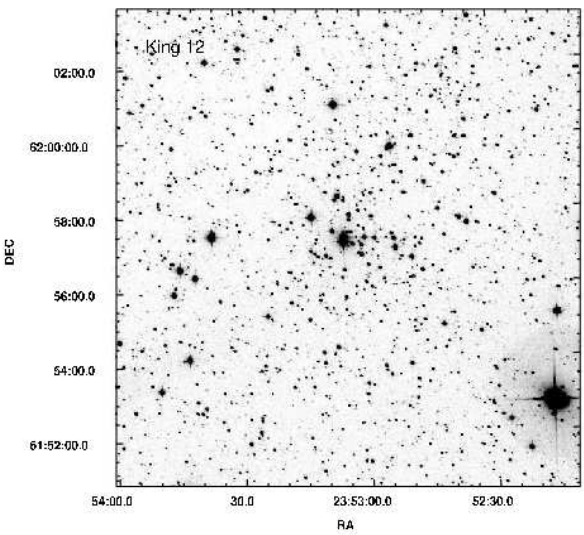}
}
\caption{The observed images ($V$-band)  of $\sim$13$\times$13 arcmin$^2$ for the clusters  Cz 4, Be 7, NGC 2236, 
 NGC 7226 and King 12.
RA and DEC refer to epoch J2000.}
\end{figure*}

\begin{figure*}
\includegraphics[width=10cm]{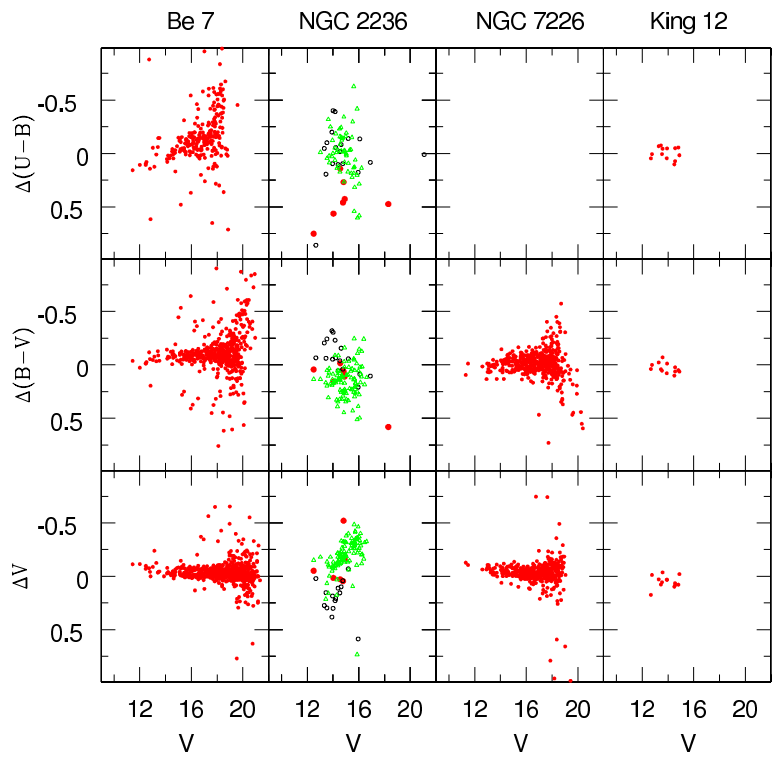}
\caption{Comparison of the present photometries and photometries available in the literature. The $\Delta$ represents present data minus data from the literature. In the case of NGC 2236 filled and open circles represent comparison with photoelectric and photographic data (Babu 1991)respectively while triangles represent comparison with data by Malisheva \& Shashkina (1987).}
\end{figure*}
\begin{figure*}
\includegraphics[width=11cm]{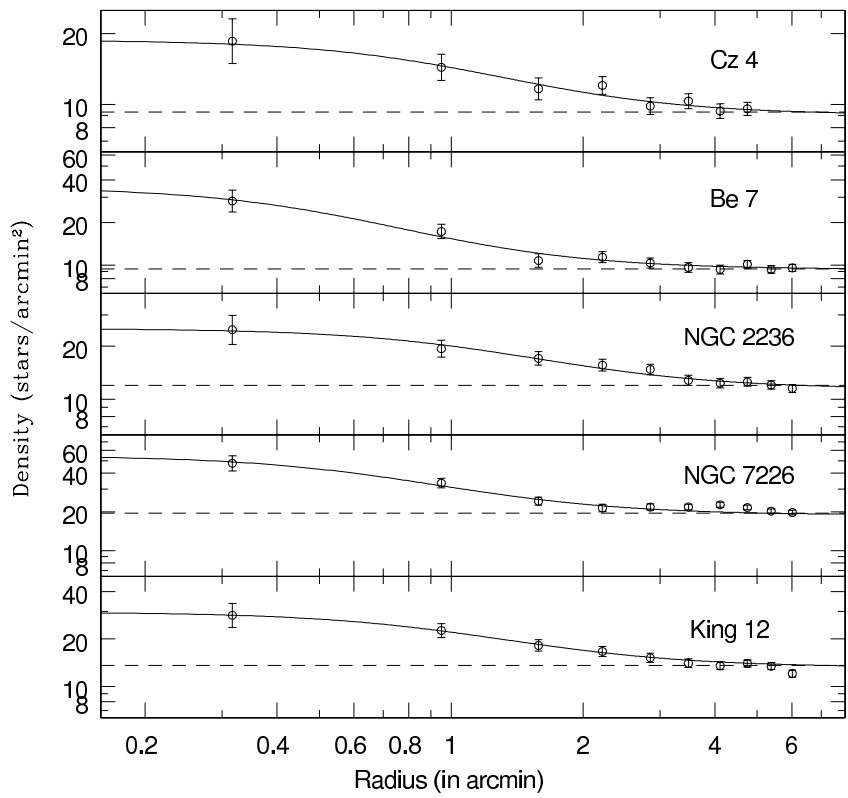}
\caption{Radial density profile for the present cluster sample. The dashed line and solid in each panel represent the field star density and King profile by Kaluzny \& Udalski (1992) respectively.}
\end{figure*}

\begin{figure*}
\includegraphics[width=14cm]{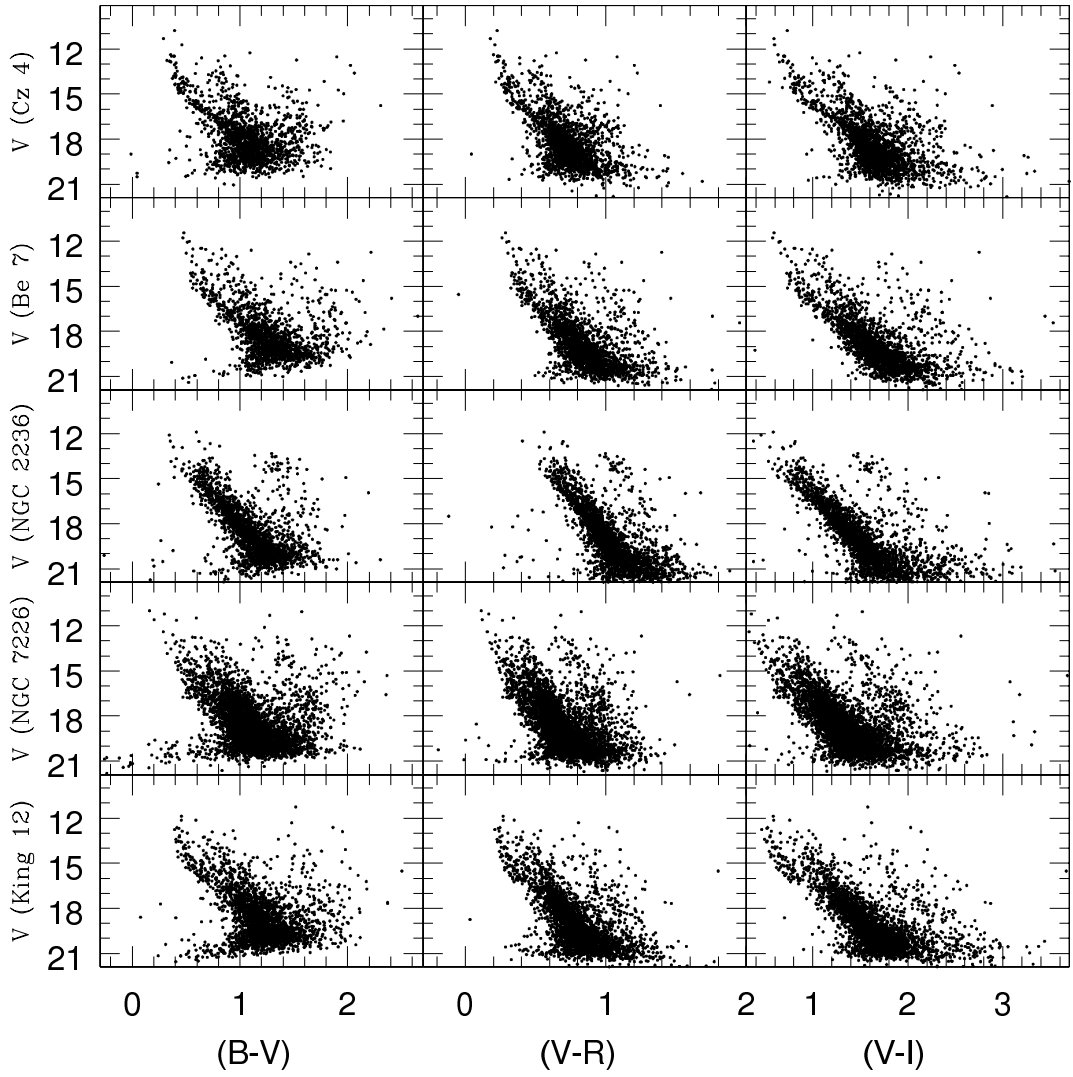}
\caption{Colour-magnitude diagrams of the present cluster sample showing all
stars detected in the frames.}
\end{figure*}

\begin{figure*}
\includegraphics[width=14cm]{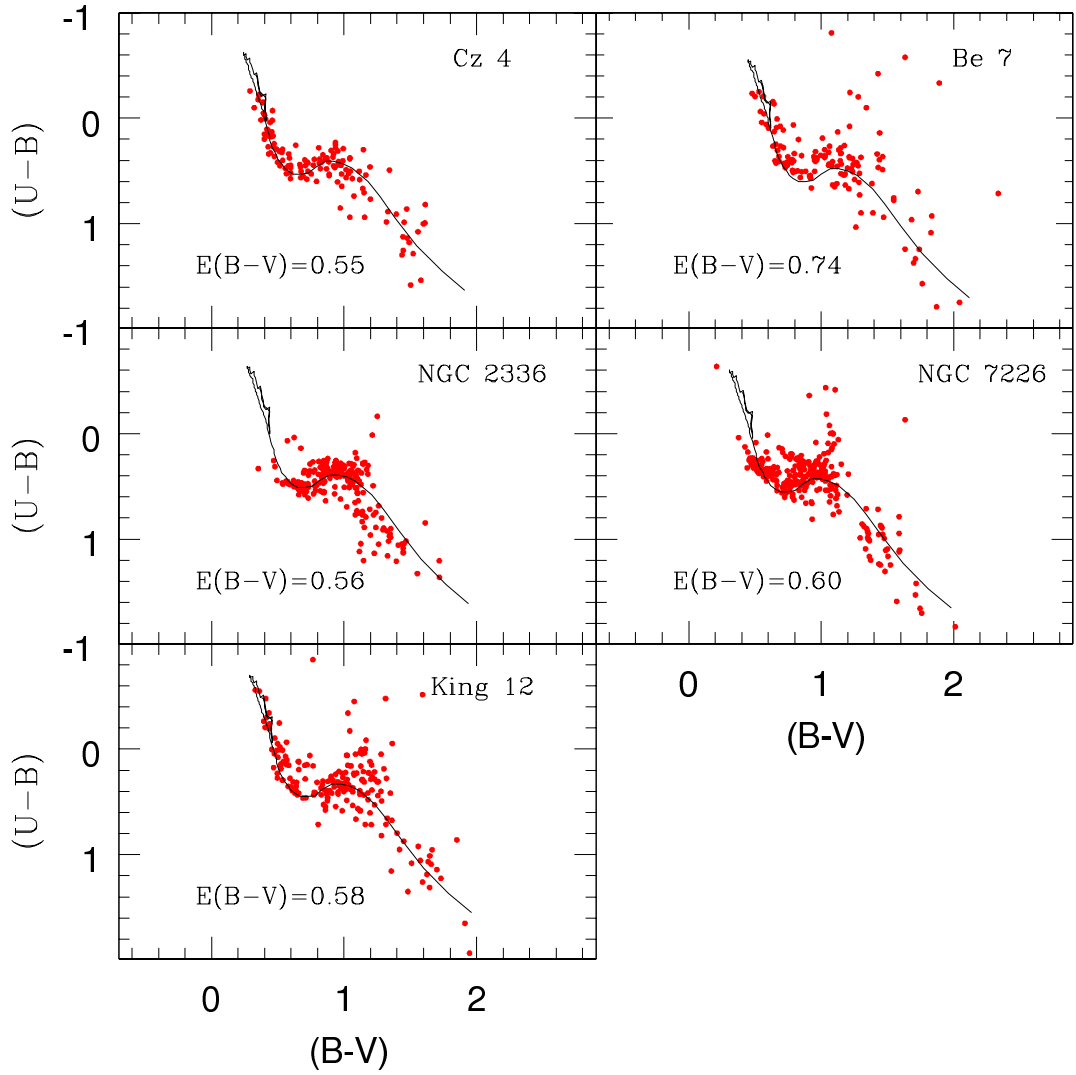}
\caption{$U-B, B-V$ colour-colour diagram of the present cluster sample. The solid line represents ZAMS by Girardi et al. (2002)
.}
\end{figure*}

\begin{figure*}
\includegraphics[width=14cm]{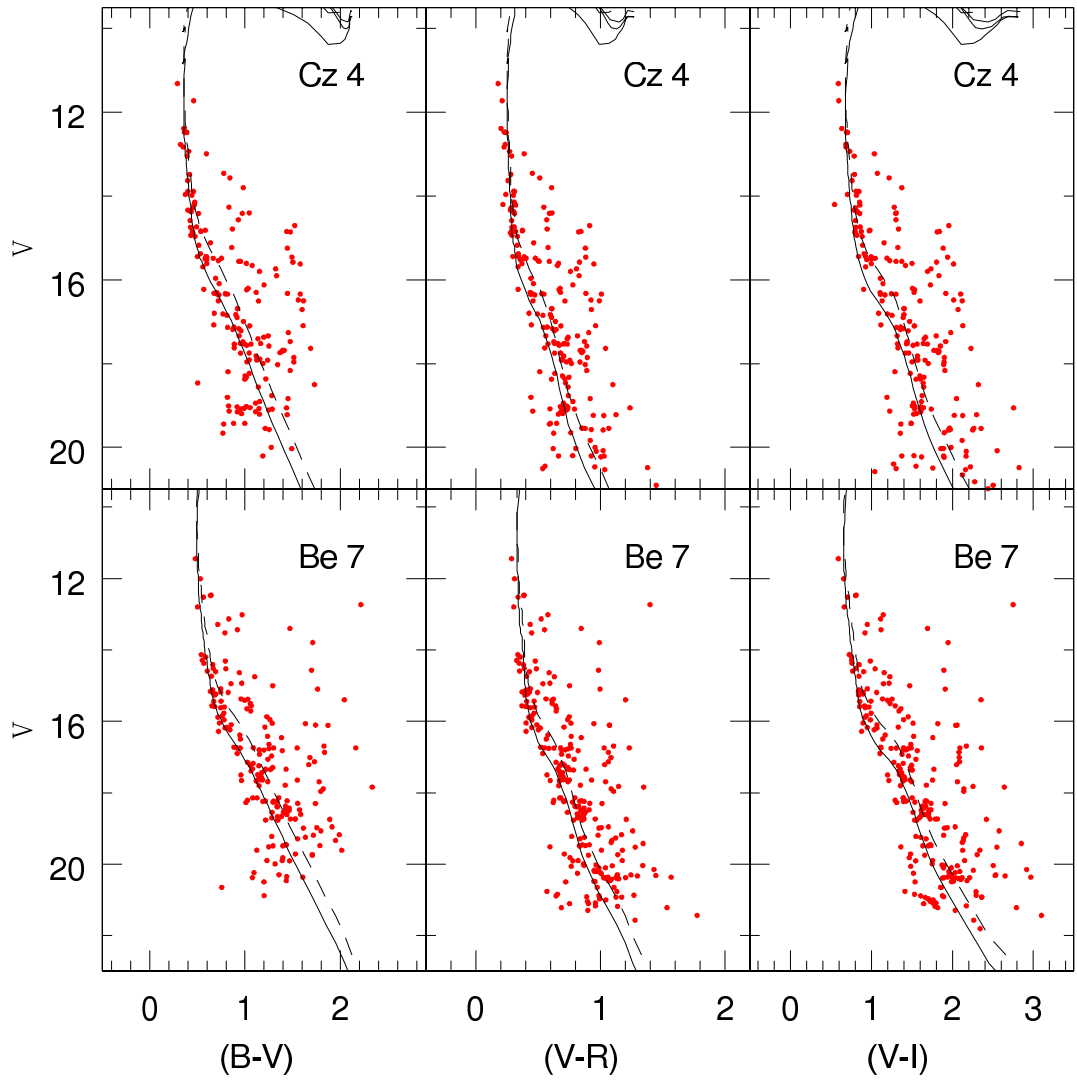}
\caption{The statically cleaned colour-magnitude diagram for Cz 4 and Be 7. The solid line represents model by Girardi et al. (2002) and dashed line is the same model but for the binary
stars
.}
\end{figure*}

\begin{figure*}
\includegraphics[width=14cm]{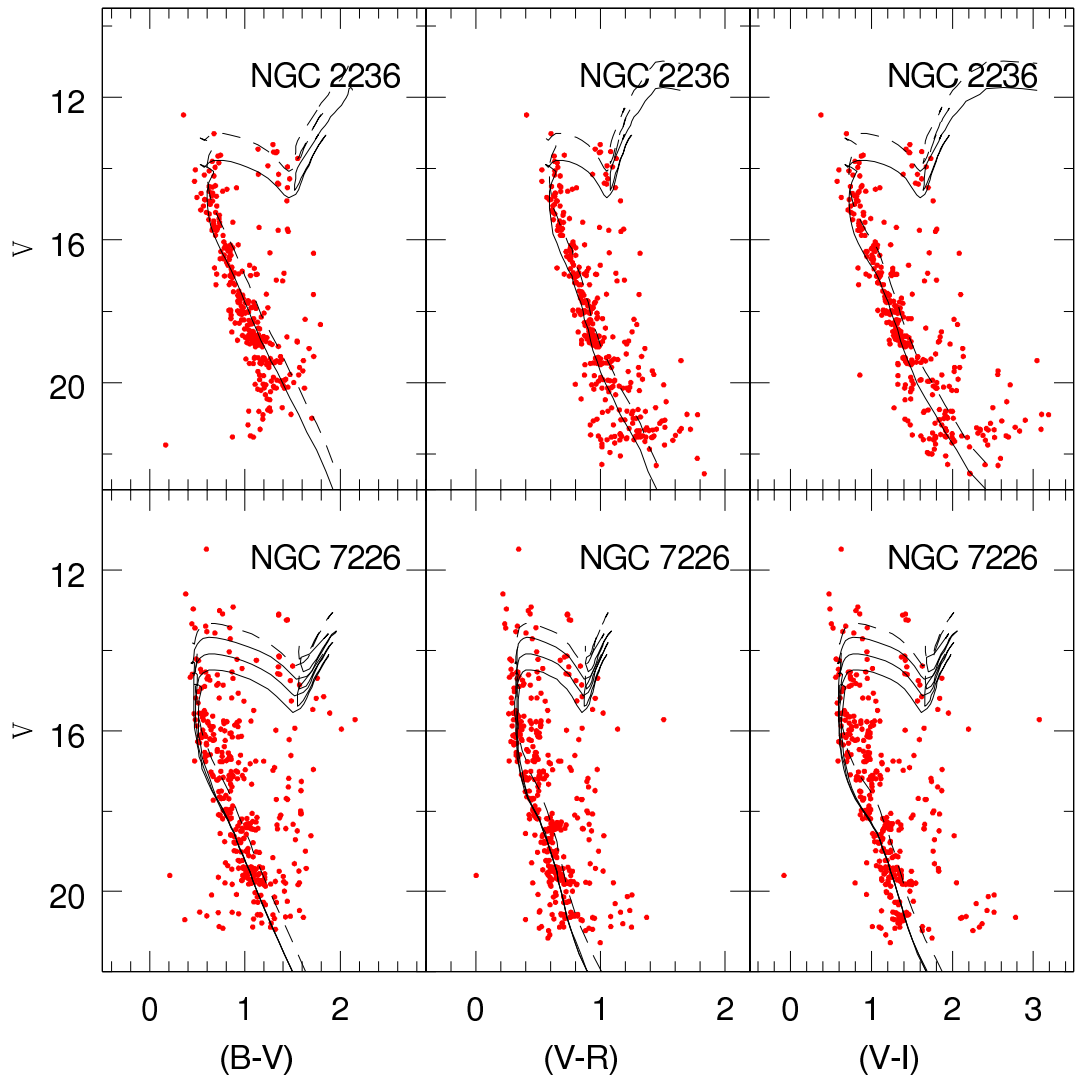}
\caption{The statically cleaned colour-magnitude diagram for King 12 and NGC 2236 NGC 7226. The solid line represents model by Girardi et al. (2002) and dashed line is the same model but for the binary
stars. In case of NGC 7226 the solid line represents model by Girardi et al. (2002) for log (age)=8.3, 8.4 and 8.5 
 and dashed line is the model of log (age)=8.4 by Girardi et al. (2002) but for the binary
stars.}
\end{figure*}

\begin{figure*}
\includegraphics[width=14cm]{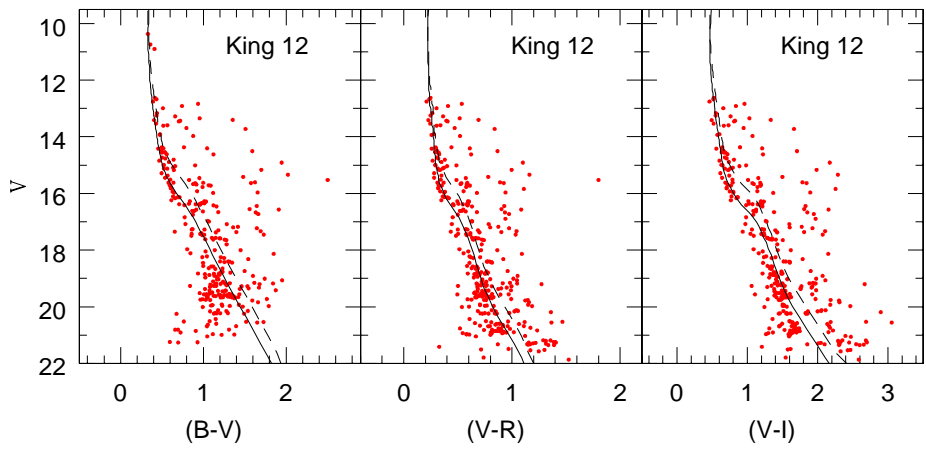}
\caption{The statically cleaned colour-magnitude diagram for King 12.  Data of three brighter stars have been taken from Mohan \& Pandey (1981). The solid line represents model by Girardi et al. (2002) and dashed line is the same model but for the binary
stars
.}
\end{figure*}

\begin{figure*}
\includegraphics[width=10cm]{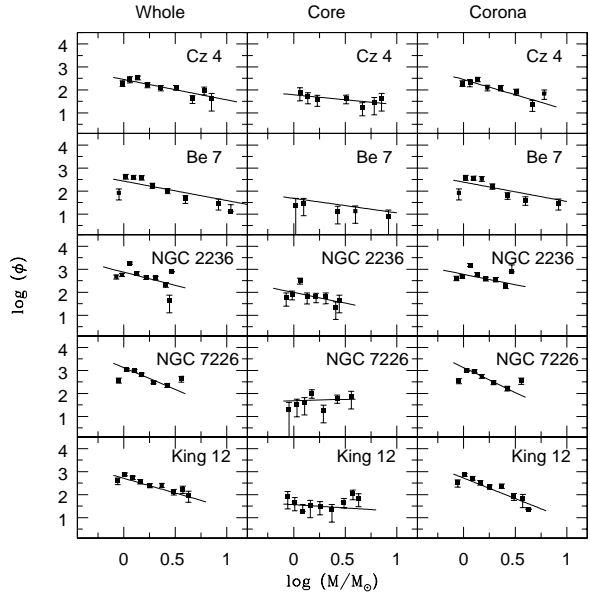}
\caption{ The cluster mass  functions for two sub-regions and whole cluster region. Log ($\phi$) represents log $(dN/d$log$M)$. The error bars represents $\pm$$\sqrt{N}$ errors. Continuous lines show the a least squares fit to the data.}
\end{figure*}

\begin{figure*}
\includegraphics[width=10cm]{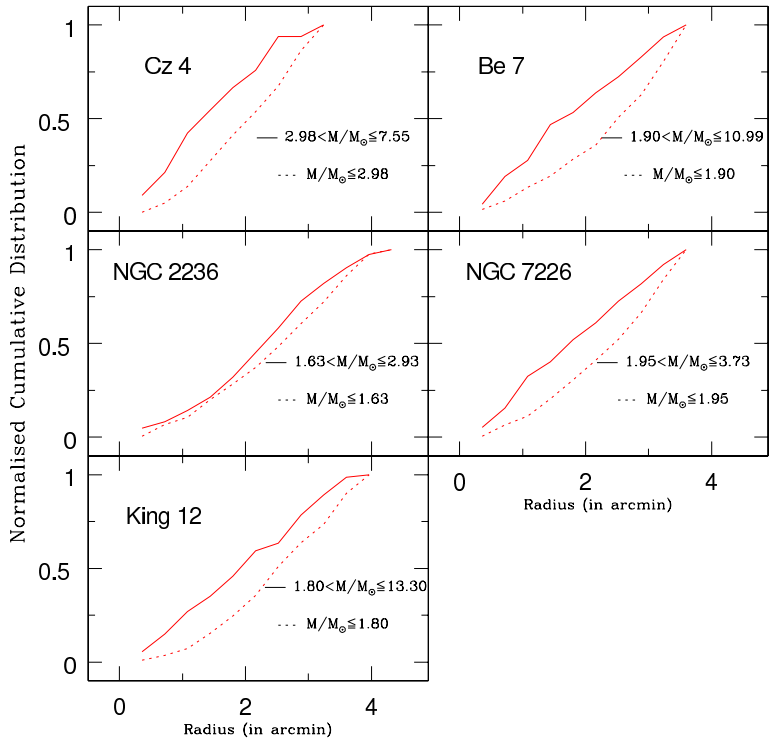}
\caption{ The cumulative radial
distribution of stars in different mass ranges.}
\end{figure*}

\begin{figure*}
\includegraphics[width=10cm]{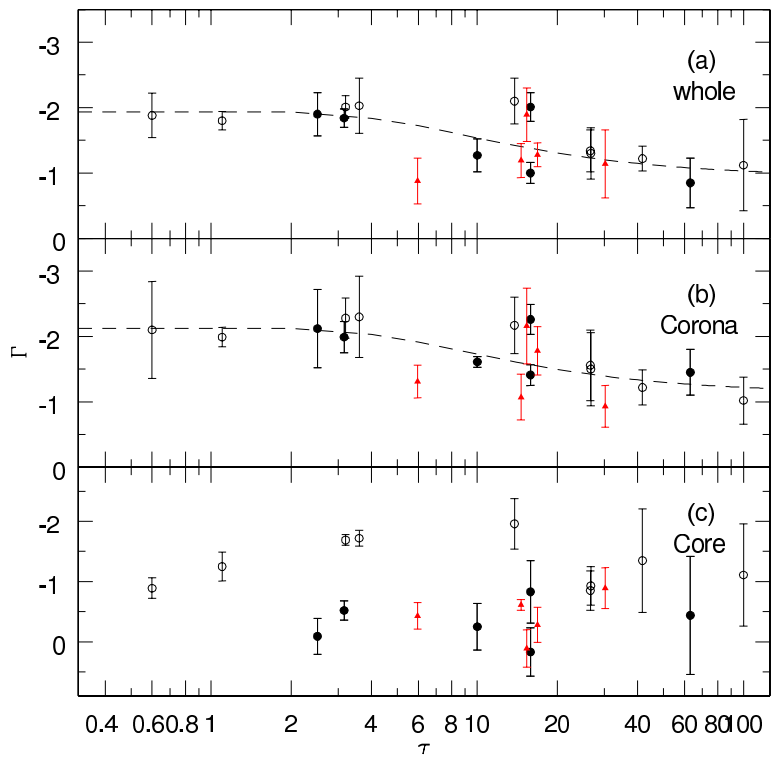}
\caption{Variation of $\Gamma$ as a function of  $\tau$ ($age/T_{E}$) for the whole cluster, corona and core regions of the clusters. Triangles, filled and open circles represent data from present study, Lata et al. (2010) and Sharma et al. (2008) respectively.
The curve represents a function $\Gamma=\Gamma_{0}+exp(a/\tau)$.}
\end{figure*}
\begin{figure*}
\includegraphics[width=12cm]{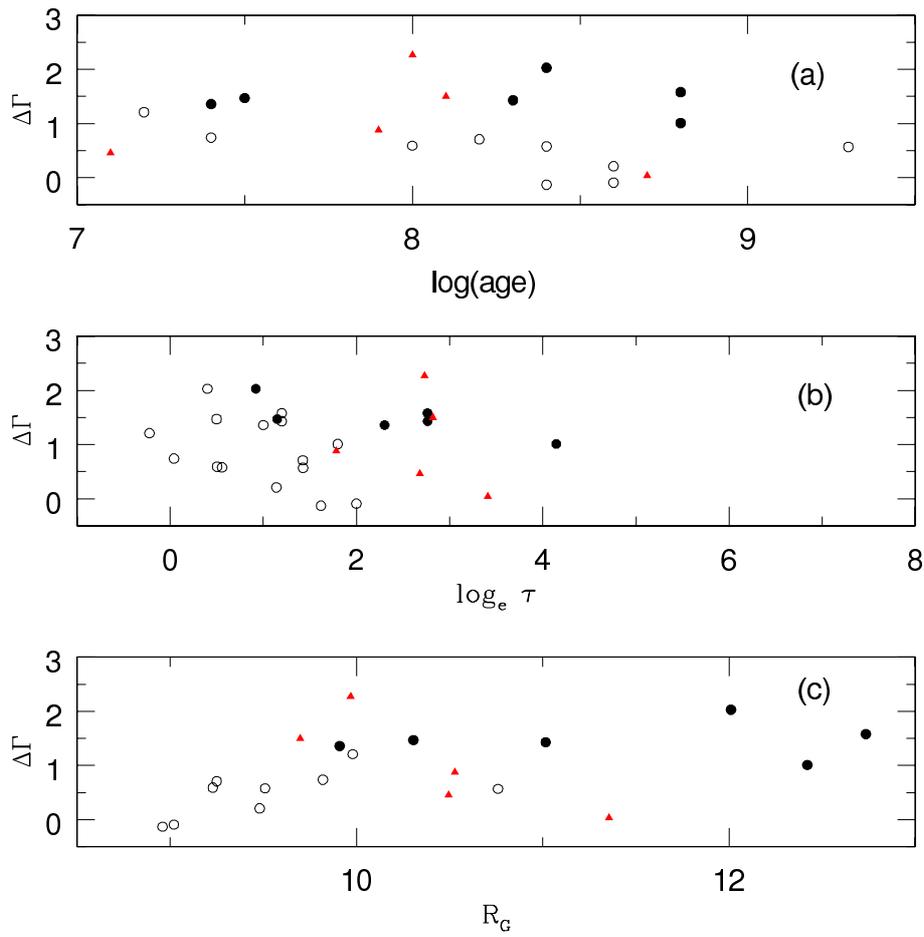}
\caption{ Variation of $\Delta \Gamma$ as function a of log (age), $\tau$ and Galactocentric distance ($R_{G}$). The symbols are same as in Fig. 11.}
\end{figure*}

%Tables from here
%Table 1------------------------------------------------------------------
\clearpage
\begin{table*}
\centering
\caption{Basic Parameters of the target clusters taken from the WEBDA.}
\begin{tabular}{llccc}
\hline
Cluster&RA (2000)&Dec (2000) &l &b   \\
&       (hh:mm:ss)&($^o$:$^\prime$:$^\prime$$^\prime$)& ($^o$)& ($^o$)   \\
\hline
Cz 4    &01 35 24& +61 16 00& 128.179&-0.945 \\
Be 7    &01 54 12 &+62 22 00&130.138&0.376 \\
NGC 2236&06 29 39&+06 49 48&204.370&-1.699 \\
NGC 7226&22 10 26&+55 23 54&101.405& -0.596\\
King 12 &23 53 00&+61 58 00&116.124&-0.130 \\
\hline
\end{tabular}
\end{table*}

%--------------------------------------------------------------------------
\begin{table*}
\centering
\scriptsize
\caption{Log of  CCD observations.}
\begin{tabular}{p{1.4cm}p{6.8cm}p{2.3cm}}
\hline
\label{logobs}
Region &  Filters and Exposure time$\times$no. of frame & Date \\
&  ~~~~~~~~~~~~~~~~(in seconds) & \\
\hline
Cz 4  & U 1200$\times$2; B 900$\times$3; V 500$\times$3; R 200$\times$3; I 200$\times$3 & 24 Dec 2008  \\
&       U 300$\times$3; B 150$\times$3; V 30$\times$3; R 30$\times$5; I 30$\times$4& 08 Dec 2007     \\
SA 98&       U 300$\times$7; B 100$\times$7; V 60$\times$7; R 60$\times$7; I 60$\times$7& 08 Dec 2007     \\
&&\\

Be 7& U 300$\times$3; B 180$\times$3; V  120$\times$3; R 60$\times$3; I 60$\times$3 & 23 Jan 2006  \\
SA 98& U 300$\times$5; B 180$\times$5; V  120$\times$5; R 60$\times$5; I 60$\times$5 & 23 Jan 2006  \\
&& \\

NGC 2236& U 1500$\times$2; B 900$\times$3; V 500$\times$3; R 300$\times$3; I 300$\times$3 & 05 Dec 2007 \\
&         U 300$\times$2; B 120$\times$2; V 90$\times$2; R 60$\times$2; I 60 $\times$2 &20 Feb 2009 \\
SA 98&         U 300$\times$2; B 150$\times$2; V 90$\times$2; R 60$\times$2; I 60 $\times$2 & 20 Feb 2009 \\
&& \\

NGC 7226& U 1200$\times$2; B 600$\times$2; V 500$\times$4; R 200$\times$3; I 200$\times$3 &11 Nov 2007 \\
&         U 300$\times$2; B 100$\times$2; V 90$\times$3; R 50$\times$3; I 40 $\times$3 &11 Nov 2007 \\
SA 98&         U 300$\times$9; B 180$\times$9; V 120$\times$9; R 60$\times$9; I 60 $\times$9 &11 Nov 2007 \\
&& \\

King 12& U 300$\times$2; B 180$\times$3; V 120$\times$3; R 60$\times$3; I 60$\times$3 & 04 Nov 2005   \\
SA 98& U 300$\times$10; B 120$\times$10; V 60$\times$10; R 30$\times$10; I 30$\times$10 & 04 Nov 2005   \\
\hline
\end{tabular}
\end{table*}

%-------------------------------------------------------------------
\begin{table*}
\centering
\caption{The zero point constants, colour coefficients and extinction coefficients on different nights.}
\scriptsize
\begin{tabular}{lccccc}
\hline
& \multicolumn{5}{c}{Date}\\ \cline{2-6}
Parameters& 04 Nov 2005& 23 January 2006&11 Nov 2007& 08 Dec 2007 &20 Feb 2009 \\
\hline
Zero point & & & \\
constant & & & \\
$q_{1}$&4.17$\pm$0.01&4.01$\pm$0.01&4.06$\pm$0.01&4.62$\pm$0.01&5.01$\pm$0.01 \\
$q_{2}$&4.61$\pm$0.01&4.45$\pm$0.01&5.05$\pm$0.01&5.07$\pm$0.02&5.70$\pm$0.02 \\
$q_{3}$&4.60$\pm$0.01&4.47$\pm$0.01&4.97$\pm$0.01&5.19$\pm$0.02&5.36$\pm$0.01 \\
$q_{4}$&4.07$\pm$0.01&3.94$\pm$0.01&4.50$\pm$0.01&4.52$\pm$0.01&5.16$\pm$0.01 \\
$q_{5}$&6.79$\pm$0.01&6.62$\pm$0.01&7.27$\pm$0.01&7.18$\pm$0.02&7.69$\pm$0.01 \\
Colour & & & \\ 
coefficient& & & \\
$p_{1}$&$-0.04\pm$0.01&$ -0.02\pm$0.01 &$-0.02\pm$0.01&$-0.02\pm0.01$& -0.05$\pm$0.01 \\
$p_{2}$&$-0.04\pm$0.01&$ -0.03\pm$0.01 &$-0.02\pm$0.01&$-0.02\pm0.02$& -0.02$\pm$0.02 \\
$p_{3}$&$-0.05\pm$0.01&$ -0.05\pm$0.01 &$-0.05\pm$0.01&$-0.05\pm0.02$& -0.05$\pm$0.02 \\
$p_{4}$&$-0.04\pm$0.01&$ -0.01\pm$0.01 &$-0.02\pm$0.01&$-0.03\pm0.02$& -0.07$\pm$0.02 \\
$p_{5}$&$0.01\pm$0.01&$ 0.05 \pm$0.01 &$ -0.02\pm$0.01&$0.01\pm0.02$& -0.04$\pm$0.02\\
Extinction & & & \\
coefficients & & & \\
$k_{u}$&0.65$\pm$0.01&0.59$\pm$0.01&0.57$\pm$0.02&0.58$\pm$0.01&0.52$\pm$0.05 \\
$k_{b}$&0.39$\pm$0.01&0.33$\pm$0.01&0.32$\pm$0.01&0.36$\pm$0.01&0.35$\pm$0.02 \\
$k_{v}$&0.25$\pm$0.01&0.23$\pm$0.01&0.19$\pm$0.01&0.25$\pm$0.01&0.26$\pm$0.02 \\
$k_{r}$&0.19$\pm$0.01&0.16$\pm$0.01&0.13$\pm$0.01&0.15$\pm$0.01&0.15$\pm$0.01 \\
$k_{i}$&0.13$\pm$0.01&0.10$\pm$0.01&0.08$\pm$0.01&0.10$\pm$0.01&0.11$\pm$0.01 \\
\hline
\end{tabular}
\end{table*}

%-------------------------------------------------------------------
\begin{table*}
\centering
\tiny
\caption{Reddening, age and distance of the clusters.}
\begin{tabular}{lcccccccc}
\hline
Cluster&$E(B-V)$&log (age)&$(m-M)$&d& log (age)$_{lit}$& $(m-M)_{lit}$&d$_{lit}$&$E(B-V)_{lit}$   \\
&(mag)&       & (mag)&(kpc)&  & (mag)&(kpc) &(mag)     \\
\hline
Cz 4 & 0.55$\pm$0.05& 7.6$\pm$0.1& 14.1$\pm$0.1&3.01 &- &-&-&- \\
Be 7 & 0.74$\pm$0.05& 7.1$\pm$0.1& 14.4$\pm$0.1 &2.64&6.60&14.53&2.57&0.80 \\
NGC 2236&0.56$\pm$0.05&8.7$\pm$0.1&14.0$\pm$0.1&2.84&8.54&13.82&2.93&0.48 \\
NGC 7226&0.60$\pm$0.05&8.4$\pm$0.2&15.6$\pm$0.1&5.60&8.45&13.75&2.62&0.54 \\
King 12 &0.58$\pm$0.05& 7.1$\pm$0.1& 14.1$\pm$0.1&2.89&7.04&13.71&2.38&0.59 \\
\hline
\end{tabular}
\end{table*}

%-------------------------------------------------------------------
\begin{table*}
\caption{ Luminosity function of the target  clusters.
N  is the  number of  probable  cluster members  in various  magnitude
bins.}
\tiny
\begin{tabular}{|p{0.7cm}|p{0.4cm}p{0.1cm}p{0.6cm}|p{0.4cm}p{0.1cm}p{0.6cm}|p{0.4cm}p{0.1cm}p{0.6cm}|p{0.4cm}p{0.1cm}p{0.6cm}|p{0.4cm}p{0.1cm}p{0.6cm}|}
\hline
Range in &\multicolumn{15}{c|}{N}\\ \cline{2-16}
V mag&\multicolumn{3}{c|}{Cz 4}&\multicolumn{3}{c|}{Be 7} &\multicolumn{3}{c|} {NGC 2236} &\multicolumn{3}{c|} {NGC 7226} &\multicolumn{3}{c|} {King 12} \\ \cline{2-16}
     &whole  &core&corona& whole& core&corona&whole&core&corona&whole  &core&corona&whole  &core&corona \\ 
     &cluster&    && cluster&    &&cluster&   &&cluster&   &&cluster&   & \\
\hline
11-12 & 2  & 2&- & 1  & - &1   & -  & - &- & - & -  &-  & -  & -&- \\
12-13 & 7  & 2&5 & 4  & 1 &4   & 1  & - &1 & - & -  &-  & 4  & 3&1 \\
13-14 & 7  & 3&4 & -  & - &-   & 2  & 2 &0 & -  & - &-  &8   & 5&3 \\
14-15 & 18 & 6&12& 7  & 2 &6   & 19 & 2 &17& 12 & 2 &10 &14  & 5&9 \\
15-16 & 14 & -&14& 16 & 2 &10  & 27 & 4 &23& 26 & 7 &19 &24  & 2&22\\
16-17 & 17 & 4&13& 16 & - &15  & 35 & 5 &30& 35 & 2 &33 &24  & 3&21\\
17-18 & 27 & 4&23& 22 & - &20  & 43 & 4 &39& 38 & 6 &32 &22  & 2&20\\
18-19 & 11 & 3&8 & 28 & 2 &26  & 70 & 12&58& 49 & 2 &47 &29  & 1&28\\
19-20 & 13 & -&13& 18 & 1 &16  & 43 & 6 &37& 65 & 2 &63 &49  & 3&46\\
20-21 & 1 & -&1 &  4  & - &4   & 24 & 3 &21& 18 & 10&17 &10  & 2&8 \\
\hline
\end{tabular}
\end{table*}

%-------------------------------------------------------------------
\begin{table*}
\caption{MF slope $\Gamma$ for core, corona and whole cluster region in the given mass range.}
\scriptsize
\begin{tabular}{|l|c|ccc|}
\hline
cluster & Mass Range&\multicolumn{3}{c|} {$\Gamma$} \\ \cline{3-5}
      & $M_{\odot}$  &core & corona & whole cluster  \\ \hline
Cz 4    &0.83-7.55    &$-0.43\pm0.20$ &  $-1.31\pm0.28$ & $-0.88\pm$0.19 \\
Be 7    &0.89-10.99  &$-0.61\pm0.09$ &  $-1.07\pm0.35$ & $-1.19\pm$0.26 \\
NGC 2236&0.87-2.93   &$-0.93\pm0.32$ &  $-0.89\pm0.34$ & $-1.14\pm$0.52 \\
NGC 7226&0.90-3.73   &$ 0.11\pm0.58$ &  $-2.16\pm0.15$ & $-1.89\pm$0.20 \\
King 12 &0.88-13.30 &$-0.28\pm0.20$ &  $-1.78\pm0.23$ & $-1.28\pm$0.18 \\
\hline
\end{tabular}
\end{table*}

%-------------------------------------------------------------------
\begin{table*}
\caption{Various parameters for the target clusters in the given mass range (see Table 6) used for calculating the dynamical relaxation time $T_{E}$. }
\tiny
\begin{tabular}{p{1.3cm}p{0.7cm}p{.6cm}p{.7cm}p{1.0cm}p{0.7cm}p{1.1cm}p{0.7cm}}
\hline
Cluster & N  & $R_{h}$& Mass& $\bar{m}$& Age & $T_{E}$ &$\tau$ \\
        &      & (pc)& ($M_{\odot}$)& ($M_{\odot}$)& (Myrs) & (Myrs) &  \\
\hline
Cz 4    & 117&1.46&276$\pm$17&2.35&39.8& 6.6 &5.97 \\
Be 7   & 116&1.65&235$\pm$14&2.03 &125&8.57& 14.59\\
NGC 2236& 264&1.96&370$\pm$26&1.40&501&16.57& 30.24\\
NGC 7226& 247&2.04&386$\pm$19&1.56&251&16.35& 15.35\\
King 12 & 187&1.30&316$\pm$15&1.68&125&7.43& 16.82\\
\hline
\end{tabular}
\end{table*}

%-------------------------------------------------------------------

\end{document}